\input harvmac\skip0=\baselineskip
\divide\skip0 by 2

\lref\withol{E.~Witten,
``Anti-de Sitter space and holography,''
Adv.\ Theor.\ Math.\ Phys.\  {\bf 2}, 253 (1998).}
\lref\gkp{S.~S.~Gubser, I.~R.~Klebanov and A.~M.~Polyakov,
``Gauge theory correlators from non-critical string theory,''
Phys.\ Lett.\ B {\bf 428}, 105 (1998).}
\lref\park{M.~Park,
``Statistical entropy of three-dimensional Kerr-de Sitter space,''
Phys.\ Lett.\ B {\bf 440}, 275 (1998).}
\lref\gubb{S.~S.~Gubser,
``Non-conformal examples of AdS/CFT,''
Class.\ Quant.\ Grav.\  {\bf 17}, 1081 (2000).}
\lref\zam{A.~B.~Zamolodchikov,
``'Irreversibility' Of The Flux Of The Renormalization Group In A 2-D Field Theory,''
JETP Lett.\  {\bf 43}, 730 (1986)
Pisma Zh.\ Eksp.\ Teor.\ Fiz.\  {\bf 43}, 565 (1986).}
\lref\gub{
D.~Z.~Freedman, S.~S.~Gubser, K.~Pilch and N.~P.~Warner,
``Renormalization group flows from holography supersymmetry and a  c-theorem,''
Adv.\ Theor.\ Math.\ Phys.\  {\bf 3}, 363 (1999).}
\lref\sonn{N.~Itzhaki, J.~M.~Maldacena, J.~Sonnenschein and S.~Yankielowicz,
``Supergravity and the large N limit of theories with sixteen  supercharges,''
Phys.\ Rev.\ D {\bf 58}, 046004 (1998).}
 \lref\ascv{
A.~Strominger and C.~Vafa, ``Microscopic Origin of the
Bekenstein-Hawking Entropy,'' Phys.\ Lett.\ B {\bf 379}, 99 (1996)
[hep-th/9601029].} \lref\boussoc{ R.~Bousso, ``Holography in
general space-times,'' JHEP {\bf 9906}, 028 (1999)
[hep-th/9906022].} \lref\hulla{
C.~M.~Hull, ``Timelike T-duality, de~Sitter space, large N gauge
theories and topological field theory,'' JHEP {\bf 9807}, 021
(1998) [hep-th/9806146].} \lref\hullb{ C.~M.~Hull and R.~R.~Khuri,
``Worldvolume theories, holography, duality and time,'' Nucl.\
Phys.\ B {\bf 575}, 231 (2000) [hep-th/9911082].}
\lref\dscft{A.~Strominger, ``The dS/CFT Correspondence,''
hep-th/0106113.} \lref\jmas{ J.~Maldacena and A.~Strominger,
``Statistical entropy of de~Sitter space,'' JHEP {\bf 9802}, 014
(1998) [gr-qc/9801096].} \lref\gibbons{ G.~W.~Gibbons and
S.~W.~Hawking, ``Cosmological Event Horizons, Thermodynamics, And
Particle Creation,'' Phys.\ Rev.\ D {\bf 15}, 2738 (1977).}
\lref\wu{ F.~Lin and Y.~Wu, ``Near-horizon Virasoro symmetry and
the entropy of de Sitter space in  any dimension,'' Phys.\ Lett.\
B {\bf 453}, 222 (1999) [hep-th/9901147].} \lref\banksb{ T.~Banks,
``Cosmological breaking of supersymmetry or little Lambda goes
back to  the future. II,'' hep-th/0007146.} \lref\banados{
M.~Banados, T.~Brotz and M.~E.~Ortiz, ``Quantum three-dimensional
de~Sitter space,'' Phys.\ Rev.\ D {\bf 59}, 046002 (1999)
[hep-th/9807216].} \lref\kim{ W.~T.~Kim, ``Entropy of 2+1
dimensional de Sitter space in terms of brick wall method,''
Phys.\ Rev.\ D {\bf 59}, 047503 (1999) [hep-th/9810169].}
\lref\banks{ T.~Banks and W.~Fischler, ``M-theory observables for
cosmological space-times,'' hep-th/0102077.} 
\lref\park{M.~Park, ``Statistical entropy of three-dimensional
Kerr-de~Sitter space,'' Phys.\ Lett.\ B {\bf 440}, 275 (1998)
[hep-th/9806119].} \lref\vijay{ V.~Balasubramanian, P.~Horava and
D.~Minic, ``Deconstructing de Sitter,'' JHEP {\bf 0105}, 043
(2001) [hep-th/0103171].} 
\lref\witten{E.~Witten, ``Quantum gravity
in de Sitter space'', hep-th/0106109.}
\lref\pm{ P. O. Mazur and E. Mottola, 
   "Weyl Cohomology and the Effective Action for Conformal Anomalies," 
hep-th/0106151.} 
\lref\witc{ E.~Witten, ``Quantum gravity
in de Sitter space'' Strings 2001 online proceedings
http://theory.theory.tifr.res.in/strings/Proceedings }
\lref\park{M.~Park, ``Statistical entropy of three-dimensional
Kerr-de~Sitter space,'' Phys.\ Lett.\ B {\bf 440}, 275 (1998)
[hep-th/9806119].} \lref\cc{See e.g. S.~Perlmutter, ``Supernovae,
dark energy, and the accelerating universe: The status of  the
cosmological parameters,'' in {\it Proc. of the 19th Intl. Symp.
on Photon and Lepton Interactions at High Energy LP99 } ed. J.A.
Jaros and M.E. Peskin, Int.\ J.\ Mod.\ Phys.\ A {\bf 15S1}, 715
(2000).} \lref\bh{ J.~D.~Brown and M.~Henneaux, ``Central Charges
In The Canonical Realization Of Asymptotic Symmetries: An Example
From Three-Dimensional Gravity,'' Commun.\ Math.\ Phys.\  {\bf
104}, 207 (1986).} \lref\juan{J.~Maldacena, ``The large N limit of
superconformal field theories and supergravity,'' Adv.\ Theor.\
Math.\ Phys.\  {\bf 2}, 231 (1998)
[hep-th/9711200].}
\lref\amm{I. Antoniadis, Pawel O. Mazur, E. Mottola, 
   "Comment on "Nongaussian Isocurvature Perturbations from Inflation" ", 
astro-ph/9705200.}
\lref\cca{
B.~P.~Schmidt {\it et al.},
``The High-Z Supernova Search: Measuring Cosmic Deceleration and
Global Curvature of the Universe Using Type Ia Supernovae,''
Astrophys.\ J.\  {\bf 507}, 46 (1998).}
\lref\rs{L.~J.~Randall and R.~Sundrum,
``An alternative to compactification,''
Phys.\ Rev.\ Lett.\  {\bf 83}, 4690 (1999).}

\lref\nima{N.~Arkani-Hamed, S.~Dimopoulos and G.~R.~Dvali,
``The hierarchy problem and new dimensions at a millimeter,''
Phys.\ Lett.\ B {\bf 429}, 263 (1998).}
\lref\ccb{
A.~G.~Riess {\it et al.}  [Supernova Search Team Collaboration],
``Observational Evidence from Supernovae for an Accelerating Universe
and a Cosmological Constant,''
Astron.\ J.\  {\bf 116}, 1009 (1998).}

\lref\ccc{S.~Perlmutter {\it et al.}  [Supernova Cosmology Project Collaboration],
``Measurements of Omega and Lambda from 42 High-Redshift Supernovae,''
Astrophys.\ J.\  {\bf 517}, 565 (1999).}

\def\ds{dS$_4$}
\def\hi{$\hat { \cal I}^+$}
\def\im{$ { \cal I}^-$}
\def\ip{$ { \cal I}^+$}

\def\ch{{\cal H}}
\def\msurr{\mathsurround=0pt}
\def\overleftrightarrow#1{\vbox{\msurr\ialign{##\crcr
        $\leftrightarrow$\crcr\noalign{\kern-1pt\nointerlineskip}
        $\hfil\displaystyle{#1}\hfil$\crcr}}}
%\draft
\Title{\vbox{\baselineskip12pt\hbox{hep-th/010087}\hbox{}
\hbox{}}}{Inflation and the dS/CFT Correspondence}

\centerline{Andrew Strominger}
\bigskip\centerline{Jefferson Physical Laboratory}
\centerline{Harvard University} \centerline{Cambridge, MA 02138}
\vskip .3in \centerline{\bf Abstract} {It is speculated that the
observed universe has a dual representation as renormalization
group flow between two conformal fixed points of a
three-dimensional Euclidean field theory. The infrared fixed point
corresponds to the inflationary phase in the far past. The
ultraviolet fixed point corresponds to a de Sitter phase dominated
by the cosmological constant indicated in recent astronomical
data. The monotonic decrease of the Hubble parameter corresponds
to the irreversibility of renormalization group flow. }

\smallskip
\Date{}
%\listtoc
%\writetoc
%\newsec{Introduction}

Recent observations \refs{\cca \ccb -\ccc}, together with the
theory of inflation, suggest the possibility that our universe
approaches  de Sitter geometries in both the far past and the far
future, but with values of the cosmological constant that differ
by a hundred or so orders of magnitude.  In recent theoretical
work \dscft , it was conjectured that a fully quantum theory,
including gravity, in pure de Sitter space with a fixed
cosmological constant has a certain dual representation as a
conformally invariant Euclidean field theory on the boundary of de
Sitter space.\foot{This conjecture followed related observations
in \refs{\amm \jmas \park \hulla \banados \kim \wu \boussoc  \hullb
\banksb
 \witc  \banks -\vijay  }, and was inspired by  an analogy to related
 early \bh\ and modern  \refs{ \juan
\gkp-\withol  } results obtained in AdS (anti-de Sitter space) . 
See also \refs{\witten,\pm}. }
The purpose of this short note is to extend this ``dS/CFT
correspondence'' so as to potentially include our own universe.

We begin with a brief synopsis of the relevant portions of \dscft\
for the case of four dimensional de Sitter space,
\ds.\foot{Reference \dscft\ focused on the case of three spacetime
dimensions because of the extra control provided by the
infinite-dimensional enhancement of the conformal group for that
case.} We assume that all observables can be generated from the
complete set of quantum correlation functions whose arguments lie
on the asymptotic boundary of \ds.  For a cosmology like our own
the relevant boundary is \hi, the Euclidean $R^3$ at future
infinity.\foot{For a discussion of other boundaries see \dscft,
and footnote 4 below.} The correlators on \hi\ of course must
transform covariantly under general coordinate transformations. Of
particular interest are the $SO(4,1)$ coordinate transformations
which are isometries of \ds. It turns out that these act on \hi\
as the $SO(4,1)$ conformal group of Euclidean $R^3$.  The
conjecture is that the \hi\ correlators are generated by a
three-dimensional conformal field theory whose conformal group is
identified with the \ds\ isometry group . That is, there are two
dual representations of the same theory, one as a bulk quantum
theory of gravity on \ds , and one as a purely spatial conformal
field theory without gravity on $R^3$.

The conjecture was motivated in \dscft\ by an analysis of the
so-called asymptotic symmetry group of de Sitter space and its
generators, together with an appropriately crafted analogy to the
by now well-established AdS/CFT correspondence \refs{\bh \juan
\gkp -\withol}. A primary difficulty with the dS/CFT conjecture is
the absence of any well-controlled example of a de Sitter solution
to string theory (or other form of quantum gravity) in which
concrete calculations are possible and the conjecture can be
tested. There is no direct construction of the boundary field
theory, although some general characteristics can be indirectly
deduced such as the relation between conformal weights and
particle masses and the absence of unitarity. We hope this
situation will be remedied in the near future perhaps by the
construction of appropriate de Sitter solutions to string theory.
Meanwhile, the proposal resonates well with other ideas from
string theory and seems worth pursuing, which we shall now proceed
to do without further apology.

An intriguing feature of the dS/CFT correspondence is the
identification of time evolution in the bulk with scale
transformations in the boundary. In flat Robertson-Walker
coordinates the \ds\ metric is \eqn\rw{ds^2=-dt^2+e^{2Ht}d \vec
x^2,} where $H$ is the Hubble constant. This geometry is isometric
under the modified time translations \eqn\frt{  t\to t+\lambda,
~~~~~~~~~\vec x \to e^{-\lambda H} \vec x,} whose generator we
will refer to as $\ch$. From the first term in \frt\ we see that
$\ch$ generates time evolution in the bulk gravity theory, while
from the second term we see that it generates scale
transformations in the boundary theory. Furthermore late times in
the bulk correspond to the UV (ultraviolet) of the boundary field
theory, while early times correspond to the IR (infrared).

Let us now assume that our universe is well-approximated by a
geometry of the  Robertson-Walker
form \eqn\rw{ds^2=-dt^2+R^2(t)d \vec x^2,} where
at early times \eqn\tin{t\to -\infty,~~~~~{\dot R \over R}\to
H_i,} while at late times  \eqn\tin{t\to \infty,~~~~~{\dot R \over
R}\to H_f .} $H_i$ here is the inflationary era Hubble constant,
typically taken to be of order $10^{24}$ cm$^{-1}$. $H_f$ is the final
value of the Hubble constant which recent observations indicate
may be of order $10^{-28}$ cm$^{-1}$.  At intermediate times $R(t)$
corresponds to standard big bang cosmology.

For such a function $R(t)$, the universe has no isometry of the
form \frt, and there would be no reason to expect a dual
representation of the bulk gravity theory as a boundary conformal
field theory. In order to interpret this, we again take our cue
from parallel developments in the study of AdS
\refs{\sonn \gub -\gubb}. The absence of a bulk isometry is conjectured
to correspond to a boundary field theory which is not conformally
invariant. Bulk time evolution is dual to RG (renormalization
group) flow in the boundary field theory. Since the isometry \frt\
is recovered for $t\to \pm \infty$, the RG flow begins at a UV
(ultraviolet) conformally invariant fixed point and ends at an IR
(infrared) conformally invariant fixed point. We note that since
late (early) times corresponds to the UV (IR)  RG flow corresponds
to evolution back in time from the future to the past.\foot{One
might have expected a discussion based on the global picture of de
Sitter space as a contracting/expanding sphere, and the associated
global time which runs from $-\infty$ on \im\ to $\infty$ on \ip.
There are several problems with this. The first is that this
picture is time reversal invariant and hence at odds with the
irreversibility of RG flow. Secondly, and more importantly,
evolution along this global time is not part of any of the de
Sitter isometries. Hence there is no way to associate it to a
conformal transformation of the boundary field theory.}

To summarize so far, it is conjectured that our universe is an RG
flow between two conformal fixed points. Time evolution is inverse
RG flow.

The Hubble parameter \eqn\hpm{H(t)={\dot R \over R}} plays a
special role in the boundary theory. To understand this we first
consider the case of dS$_3$. It was shown in \dscft\ (using
properties of the Virasoro algebra and an analysis of the
asymptotic symmetry generators) that the central charge of the
two-dimensional boundary theory obeys \eqn\ftlk{c_2={3 \over 2
HG_3},} where $G_3$ is the three-dimensional bulk Newton constant.
$c_2$ is a measure of a number of degrees of freedom of the theory
and is related to the two point function of the stress tensor. For
a generic (non-conformal) two-dimensional field theory a
``$c$-theorem'' has been proven by Zamolodchikov: an appropriately
defined $c_2$ always decreases from the UV to the IR along RG
flows \zam.\foot{The $c_2$ defined in \zam\ is not exactly the
same as that defined by \ftlk, but they agree at the
conformally-invariant fixed points.} Intuitively this corresponds
to the fact that degrees of freedom are integrated out along RG
flows and hence $c$ should decrease. Indeed an appropriate
combination of Einstein's equations can be written in the form
\eqn\acm{\dot H=-{8\pi \over 3} G_3(p+\rho).} If the null energy
condition is obeyed, the right-hand side is non-positive and $H$
decreases with time. $c_2$, as defined by \ftlk\ with a general
time-dependent $H$,  will then decrease along RG flows (which go
from the future to the past).

A similar discussion, parallel to that given for AdS$_d$
\refs{\sonn \gub -\gubb}, can be made for higher $d$-dimensional
de Sitter spaces and yields $c_{d-1} \sim {1 \over H^{d-2}G_{d}}$.
The Einstein equation then implies that this decreases along RG
flow, interpreted as inverse time evolution. In higher dimensions,
despite numerous attempts,  there is no general proof from field
theory that $c_{d-1}$ decreases along RG flow. Nevertheless, one
expects $c_{d-1}$ should be a measure of the number of degrees of
freedom of the boundary theory.\foot{However we note that, 
in the present 
context, this expectation is clouded by the fact that the 
boundary field theories are probably not unitary \dscft.} Indeed for $dS_4$ an appropriate
combination of Einsteins equations can be written in the form
\eqn\acm{\dot H=-4\pi G(p+\rho).} If the null energy condition is
obeyed, the right hand side is non-positive, $H$ decreases with
time and $c_3$ decreases along RG flows. 

One of the puzzling features of an expanding universe is that the
number of degrees of freedom  seems to increase with time. If
string theory or an alternative provide a cutoff at the Planck
length, then the number of degrees of freedom of the universe at
any moment of time is naively of order the spatial volume in
Planck units.\foot{The generalized second law/holographic
principle may highly constrain the possibilities for exciting many
of these degrees of freedom at the same time.} This increase in the number
of degrees of freedom seems hard to reconcile with the existence of a
unitary Hamiltonian. In the present
proposal, the extra degrees of freedom arise because time
evolution to the future is inverse RG flow, and hence corresponds
to integrating in new degrees of freedom.

A fundamental puzzle in modern cosmology is the large hierarchy
characterized by the large ratio ${H_i \over H_f}\sim 10^{52}$.
In the present proposal this turns into the field theory problem
of understanding the large ratios of the numbers of degrees of
freedom of the UV and IR fixed points. This is in a sense the
opposite of the recent discoveries of \refs{\nima,\rs} that field
theory hierarchies could be turned into problems in geometry.

The picture proposed here recasts the question ``What is the
origin of the universe?'' in a new light.  In a sense it puts the
origin of the universe in the infinite future, rather than the
infinite past, because only in the ultraviolet can all the degrees
of freedom comprising the universe be seen. The past is an IR
fixed point at which most of the degrees of freedom are not
present.  Indeed one could imagine a preinflationary stage with no
degrees of freedom at all, corresponding to a dual field theory at
\ip\ with a mass gap and trivial IR fixed point. In our picture
the universe is being continually created with the passage of
time, and come into full existence only in the infinite future.

In conclusion, recent indications that our universe is
asymptotically \ds\ in both the past and future are tailor made
for a new cosmological paradigm of the universe as RG flow.

{\bf Acknowledgments.} It is a pleasure to thank Nima Arkani-Hamed,
R. Bousso and Shiraz Minwalla
for useful discussions. This work was supported in part by DOE
grant DE-FG02-91ER40655.

\listrefs
\end